%% file: Main.tex
\documentclass{article}

\usepackage{arxiv}

\usepackage[utf8]{inputenc} 
\usepackage[T1]{fontenc}    
\usepackage{hyperref}       
\usepackage{url}            
\usepackage{booktabs}       
\usepackage{amsfonts}       
\usepackage{nicefrac}       
\usepackage{microtype}      
\usepackage{lipsum}		
\usepackage[pdftex]{graphicx}
\usepackage{natbib}
\setcitestyle{numbers}
\usepackage{doi}
\usepackage{amsmath,amssymb,amsfonts}
\usepackage{algorithmic}
\usepackage{tabularx}
\usepackage{booktabs}
\usepackage{siunitx}
\usepackage{multicol}
\usepackage{textcomp}

\usepackage{xurl}
\def\BibTeX{{\rm B\kern-.05em{\sc i\kern-.025em b}\kern-.08em
    T\kern-.1667em\lower.7ex\hbox{E}\kern-.125emX}}
\usepackage[nolist]{acronym}
\def\BibTeX{{\rm B\kern-.05em{\sc i\kern-.025em b}\kern-.08em
    T\kern-.1667em\lower.7ex\hbox{E}\kern-.125emX}}
\usepackage{url}
\hypersetup{ breaklinks=true, 
colorlinks=false}

\begin{document}
\title{COVID-19 vaccination certificates in the Darkweb}


\author{{\hspace{1mm}Dimitrios Georgoulias} \\
	Cyber Security Group\\
	Aalborg University\\
	Copenhagen, 2450 \\
	\texttt{dge@es.aau.dk} \\
	\And
	Jens Myrup Pedersen\\
	Cyber Security Group\\
	Aalborg University\\
	Copenhagen, 2450 \\
	\texttt{jens@es.aau.dk} \\
	\And
    Morten Falch \\
	Cyber Security Group\\
	Aalborg University\\
	Copenhagen, 2450 \\
	\texttt{falch@es.aau.dk} \\
	\And
    Emmanouil Vasilomanolakis \\
	Cyber Security Group\\
	Aalborg University\\
	Copenhagen, 2450 \\
	\texttt{emv@es.aau.dk} \\
}


\renewcommand{\shorttitle}{COVID-19 vaccination certificates in the Darkweb}

\hypersetup{
pdftitle={A template for the arxiv style},
pdfsubject={q-bio.NC, q-bio.QM},
pdfauthor={David S.~Hippocampus, Elias D.~Striatum},
pdfkeywords={First keyword, Second keyword, More},
}

\begin{acronym}
    \acro{rnd}[R\&D]{Research and Development}
    \acro{PPI}{Pay-per-Install}
    \acro{PPC}{Pay-per-Click}
    \acro{P2P}{Peer-to-Peer}
    \acro{FE}{Finalize Early}
    \acro{FAQ}{Frequently Asked Questions}
    \acro{AVCs}{Automated Vending Carts}
    \acro{AVC}{Automated Vending Cart}
    \acro{OPSEC}{Operational Security}
    \acro{GPS}{Global Positioning System}
    \acro{MBBs}{Moisture Barrier Bags}
    \acro{MBB}{Moisture Barrier Bag}
    \acro{OTR}{Off-The-Record}
    \acro{XMPP}{Extensible Messaging and Presence Protocol}
    \acro{OMEMO}{OMEMO Multi-End Message and Object Encryption}
    \acro{FAQ}{Frequently Asked Questions}
    \acro{DWM}{Dark Web Marketplaces}
    \acro{FE}{Finalize Early}
    \acro{PO}{Post Office}
    \acro{ETH}{Ethereum}
    \acro{XPR}{Ripple}
    \acro{LTC}{Litecoin}
    \acro{BCH}{Bitcoin Cash}
    \acro{DASH}{Dash}
    \acro{ZEC}{Zcash}
    \acro{XVG}{Verge}
    \acro{BTCP}{Bitcoin Private}
    \acro{ADA}{Cardano}
    \acro{KYC}{Know Your Customer}
    \acro{PGP}{Pretty Good Privacy}
    \acro{TOTP}{Time-based One-Time Password}
    \acro{QR}{Quick Response}
    \acro{2FA}{Two-Factor Authentication}
    \acro{CaaS}{Cybercrime-as-a-Service}
    \acro{BPHS}{Bulletproof Hosting Services}
    \acro{cnc}[C\&C]{Command and Control}
    \acro{VPN}{Virtual Private Network}
    \acro{IoT}{Internet of Things}
    \acro{HS}{Hidden Service}
    \acro{IRC}{Internet Chat Relay}
    \acro{LEA}{Law Enforcement Agency}
    \acro{LEAs}{Law Enforcement Agencies}
    \acro{BTC}{Bitcoin}
    \acro{OS}{Operating System}
    \acro{XMR}{Monero}
    \acro{IP}{Internet Protocol}
    \acro{CWG}{Conficker Working Group}
    \acro{ICANN}{Internet Corporation for Assigned Names and Numbers}
    \acro{NCA}{National Crime Agency}
    \acro{EC3}{European Cybercrime Centre}
    \acro{SMB}{Windows Server Message Block}
    \acro{FBI}{Federal Bureau of Investigation}
    \acro{ISP}{Internet Service Provider}
    \acro{AVC}{Automated Vending Cart}
    \acro{DDoS}{Distributed Denial of Service}
    \acro{HTTP}{Hypertext Transfer Protocol}
    \acro{DGA}{Domain Generation Algorithm}
    \acro{DNS}{Domain Name System}
    \acro{AFOSI}{Air Force Office of Special Investigations}
    \acro{LTC}{Litecoin}
    \acro{BCH}{Bitcoin Cash}
    \acro{Ethereum}{ETH}
\end{acronym}

\maketitle

\begin{abstract}
COVID-19 vaccines have been rolled out in many countries and with them a number of vaccination certificates. For instance, the EU is utilizing a digital certificate in the form of a QR-code that is digitally signed and can be easily validated throughout all EU countries.
In this paper, we investigate the current state of the COVID-19 vaccination certificate market in the darkweb with a focus on the EU Digital Green Certificate (DGC). We investigate $17$ marketplaces and $10$ vendor shops, that include vaccination certificates in their listings. Our results suggest that a multitude of sellers in both types of platforms are advertising selling capabilities. According to their claims, it is possible to buy fake vaccination certificates issued in most countries worldwide.
We demonstrate some examples of such sellers, including how they advertise their capabilities, and the methods they claim to be using to provide their services. We highlight two particular cases of vendor shops, with one of them showing an elevated degree of professionalism, showcasing forged valid certificates, the validity of which we verify using two different national mobile COVID-19 applications.
\end{abstract}

\input{Chapters/1.Introduction}

\input{Chapters/2.Methodology}
\input{Chapters/3.Background_and_related_work}

\input{Chapters/4.Certificates}

\input{Chapters/5.Valid}
\input{Chapters/6.Conclusion}

\bibliographystyle{plain}
\bibliography{references}  






\end{document}

%% file: Chapters/1.Introduction.tex
\section{Introduction}
The darkweb has been actively serving as a platform where cybercriminals can carry out their operations, since the founding of the \textit{Farmer's Market} (2010) \cite{farmer} and \textit{Silk Road} (2011) \cite{christin2013traveling}. Both of these marketplaces, operated via \textit{Tor} hidden services, which is still the most popular anonymization network to this day. While these marketplaces started off with a heavy focus on drugs, though the years such platforms have evolved, providing a large variety of products and services (e.g. firearms, botnet services, malware, stolen bank credentials). 

The COVID-19 pandemic has had a great impact on millions of people around the globe, affecting many different aspects of their lives. In order to reverse the worldwide disruption that the virus has caused, vaccines were developed, aiming towards protecting the population and halting the spread of the virus. 

Trading platforms on the darkweb were very quick to take advantage of the pandemic situation. Vendors started offering vaccines on several marketplaces, or on their own independent vendor shops \cite{bracci2021dark}. After the vaccine development, the next step was monitoring the vaccination status of the population, which was achieved through the issuance of vaccination certificates, in physical or digital form. In several countries, not being vaccinated is bound to cause implications in people's social and work life, often excluding them from some activities, and making daily tasks harder to carry out. For example, the non-vaccinated population needs COVID tests frequently, might be denied indoor access to restaurants, or have challenges while traveling. Since vaccinations can play such an important role in certain countries, a new market has emerged on the darkweb. Marketplaces and vendor shops are currently offering both physical and digital certificates, from a variety of countries, or fake PCR test results as an alternative, with non-vaccinated people as the target client group. Individuals that do not wish to be vaccinated, but want the convenience of owning a vaccination certificate, can visit the darkweb and purchase one on a number of different platforms.

In this paper, we focus on the COVID-19 vaccination certificate darkweb market, with an emphasis on the EU Digital Green Certificate (DGC). We investigate $17$ marketplaces and $10$ vendor shops that list physical or digital proofs of vaccination as available products, with the purpose of documenting the different aspects that compose this specific type of illegal trading. This includes elements such as countries of origin, countries that the certificate is valid in, shipping, means of communication with the vendors, as well as how these items find their way to the sellers. We then demonstrate examples of such sellers and emphasize on two interesting cases in which the vendors provide valid EU digital certificates as proof of their service's legitimacy. Notably, one of the shops presents a very high degree of professionalism. We confirm the validity of these certificates, and examine their details.

The remainder of this paper is structured as follows. In Section $2$, we give an overview of the methods used to carry out our research for the purposes of this paper. Section $3$ provides background information on the issuance and the verification of vaccination certificates, and discusses the related work. In Section $4$, we dive into the mapping of the certificate market of the darkweb. Section $5$ is dedicated to investigating the legitimacy of the listings we found. Lastly, Section $6$ concludes this paper.

%% file: Chapters/2.Methodology.tex
\section{Methodology}
\label{method}
To locate these selling points we made use of darkweb search engines, such as \textit{Torch} and \textit{Ahmia}, as well as popular clearweb and darkweb introduction points like \textit{Recon}, \textit{dark.fail}, and \textit{The Hidden Wiki}. These platforms provided some of the onion addresses of the marketplaces and vendor shops we investigated for the purpose of this paper. We also searched for discussions related to COVID-19 certificates on darkweb forums, with \textit{Dread} serving as the main resource, since it is by far the most popular darkweb forum at the time of this paper (October 2021). These discussions would often include some type of vendor or selling point advertisement. 

After locating all of these resources, we navigated through each site individually, exploring their listings and properties, such as payment methods, accepted cryptocurrencies, pricing, and preferred means of communication. Additionally, we emphasized on the vendors' status on each platform, which refers to their trust level, profile activity, client feedback received, as well as their sale history, with the purpose of gaining insight on the legitimacy of the listings we managed to discover. To achieve this goal, we also utilized two different national COVID-19 mobile applications (the French "\textit{Tousanticovid}" and the Danish "\textit{Coronapas}") to scan and validate QR codes certificates that we found being showcased by the vendors as proof of legitimacy. 

Lastly, in order to explore how forging COVID-19 certificates could theoretically be achieved, we needed to firstly understand the operation of the mechanisms that are tasked with the issuance and verification of these documents. At this point we mainly focused on the official specifications of the infrastructure in place, as documented by the European Commission \cite{certtechspecs, jsonspecs, europacerts, revocation}.

%% file: Chapters/3.Background_and_related_work.tex
\section{Background and Related Work}
In this section we provide insight on the operations of the issuance and verification mechanisms, with focus on the ones utilized in Europe. Furthermore, we also present existing research that is related to our work.
  \label{background}
    \subsection{EU certificate issuance and verification}
  
       Each issuing body (e.g. public health authorities) has a digital certificate, known as \textit{Document Signing Certificates (DSCs)} \cite{cscas}, along with one key pair \cite{europacerts}, comprised of a private and public counterpart. The validity of the DSCs (which extends to the key pair) is two years, while the recommended usage period is six months \cite{revocation}. Regarding the issuance, this private key is used to digitally sign the COVID-19 certificates, to provide proof of their validity, and protect against fraudulent proofs of vaccination. The public key is used in the verification of the COVID certificates, by providing assurance on the signature's validity, since a valid signature can only have come from the corresponding private key, namely the private key of the issuing body. 
       Additionally, every country-member has one digital certificate (in some cases there are more than one) called a \textit{Certificate Signer Certificate Authority (CSCA)} \cite{cscas}, which is self-signed by the private part of a master key pair. This private key is also used to digitally sign the DSCs, providing them with validity.
            \begin{figure}
            \centering
            \includegraphics[width=0.65\textwidth]{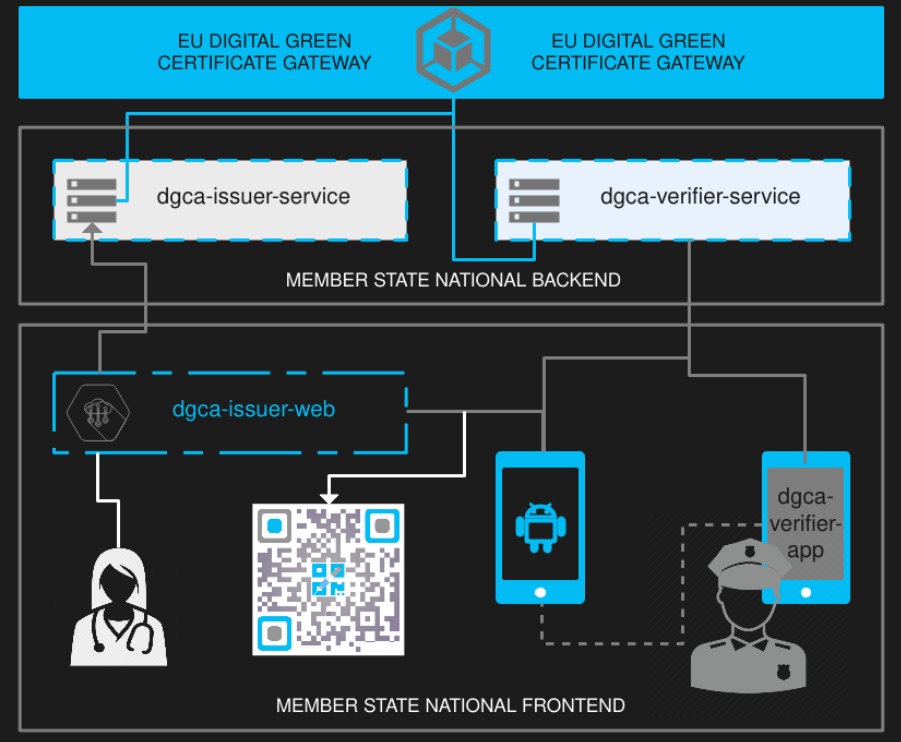}
            \vspace{-0cm}\caption{Architecture of the EU COVID-19 verification and issuance system. This Figure was discovered on one of the vendor shops and it is an adaption of the architecture Figure found in the European Commission official documentation \cite{certtechspecs}.}
            \label{fig:architecture}
            \end{figure}
                
        With the purpose of coordinating the verification of EU COVID-19 certificates all across Europe, the \textit{Digital Green Certificate Gateway (DGCG)}, developed by the European Commission, stores the public keys of every issuing body, which are then used by all of the COVID-19 mobile applications \cite{europacerts, 9558786}. More specifically, these keys are retrieved from the national \textit{issuance service} of each country, and provided to the \textit{verifier service} of each of the remaining member countries, as illustrated in Figure \ref{fig:architecture}. The list of valid keys is periodically refreshed by the mobile applications ($24$ hour interval) by connecting to the national verifier service and checking for updates, such as new or revoked certificates/keys.

\begin{figure*}[htbp!]
            \centering
            \includegraphics[width=0.75\textwidth]{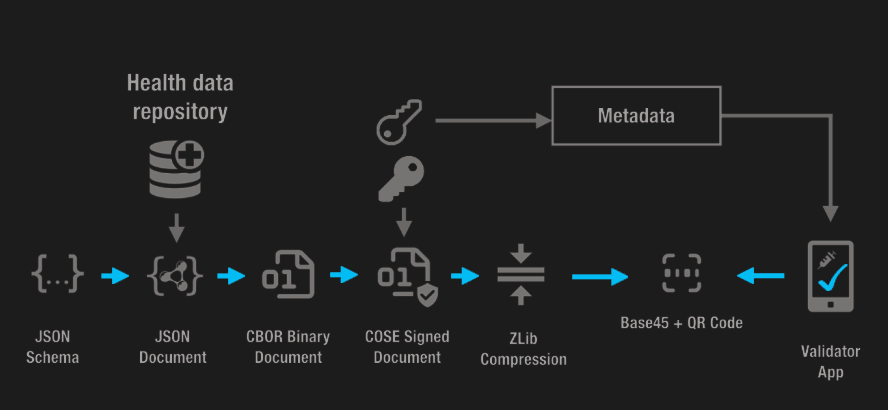}
            \vspace{-0cm}\caption{Overview of the certificate QR code generation, along with the reverse process of verification. This Figure was discovered in one of the vendor shops, and it follows the official specifications given by the European Commission \cite{certtechspecs}.}
            \label{fig:json}
            \end{figure*}

        The national certificate issuing system \cite{certtechspecs}, has two distinct components, each performing its own tasks: the issuer frontend/web application and the backend/issuer service. At the time of vaccination of an individual at a vaccination center, by using the frontend, their information is placed inside a \textit{JSON Web Token (JWT)}\footnote{The JSON includes their name, date of birth, targeted disease code, country of registration, date of vaccination, along with the issuing organization, vaccine type, manufacturer, vaccine ID code, number of registered doses, number of total doses in the series, and lastly a unique certificate ID\cite{jsonspecs}.} and encoded in the \textit{Concise Binary Object Representation (CBOR)} format \cite{cbortoken}. This information is then hashed by the frontend, and sent to the national backend, where the private keys of all issuing bodies are stored (see Figure \ref{fig:architecture}). It is then digitally signed with the private key of the issuing body that the vaccination center is operating under, using the \textit{CBOR Object Signing and Encryption (COSE)} protocol \cite{cose}. The CBOR web token along with the signature acquired by the issuer service, are both compressed at the application frontend using the \textit{zlib} library \cite{zlib}, encoded to \textit{base45}, and then the QR code is formed. The procedure is illustrated in Figure \ref{fig:json}.

        The way to verify one's certificate, is through scanning their QR code, which is available both in PDF file format, and in the mobile COVID-19 applications (in some occasions also in physical form). The verification process, is also comprised of two different segments, namely the verifier frontend/mobile app and the national backend/verifier service \cite{certtechspecs}. In the verification process all of the steps described above are reversed. After scanning the QR code using the mobile application, the base45 string is acquired, which is then decoded, decompressed, and at that point the signature can be extracted. The signature's verification is carried out using the list of trusted public keys stored in the key store of the application, after being regularly updated from the verifier service every $24$ hours, as mentioned above. Since the public keys are cached, this mechanism also allows for the offline verification of the certificates.
        
    \subsection{Certificate issuance and verification in Russia and the United States}
        In the case of the United States, it has been decided by president Joe Biden, that there will be no national mobile COVID-19 application, for all of the $50$ states. The choice has been left to the states, which individually choose whether they want to utilize a digital, physical, or any system at all, as a means to verify the population's vaccination status. At the time of this paper (October 2021), only $7$ out of the $50$ states are using a mobile application \cite{uscerts}. However, the common practice in all states is the issue of vaccination cards. At the time of vaccination, all individuals are issued a card which states the last and first name, date of birth, patient number, vaccine manufacturer, date and location of vaccination, and most importantly, the lot number. The lot number refers to the unique identifier located on each vaccine, and it serves as the method of authenticating the population's vaccinations. 
    
        Russian authorities implemented the usage of QR codes for vaccination authentication purposes in June 2021, but it was later abandoned in mid-July \cite{russiareuters}. However, it has been announced that QR codes will be utilized again from the 1st of November and onwards. According to the mechanism Russia has in place, vaccinated users can visit a specific website, input their information, and acquire a QR code file, which can be saved on their device or printed.

    \subsection{Related work}
        Since COVID-19 started affecting the cyber world equilibrium, there have been notable research efforts that attempt to shade light on pandemic-specific illegal activities. 
        
        \textit{Bracci et al.} \cite{bracci2} explore the darkweb market for COVID-19 related products, before vaccines became officially available to the public, through crawling. They find that several platforms offered products such as protective equipment (e.g. gloves and masks), tests, vaccines, and even "antidotes" and "cures". \textit{Bracci et al.} \cite{bracci2021dark}, also analyze data crawled from darkweb marketplaces after the roll out of COVID-19 vaccines, focusing on vaccine and vaccination certificate listings. Their effort resulted in documenting 5 platforms that offered vaccination certificates, with data until April 20, 2021. In our work we discover $27$ selling points, showcasing how the market has grown over the course of six months. \textit{Vu et al.} \cite{vu2020turning}, investigate how COVID-19 has affected clearweb illegal trading platforms, and specifically \textit{Hack Forums}, focusing on the economic and social aspect, as well as the reputation ecosystem that surrounds the forum's operation.
        Lastly, Karopoulos et al. surveyed how digital certificates have been implemented in the form of mobile apps \cite{9558786}.

    

%% file: Chapters/4.Certificates.tex
\section{Mapping the vaccination certificate market}
In this section we illustrate each aspect of the certificate purchasing procedure, from locating the vendors, up until the finalization of the order, by navigating through $10$ vendor shops and $17$ marketplaces on the darkweb.

\begin{figure}
            \centering
            \includegraphics[width=0.99\columnwidth]{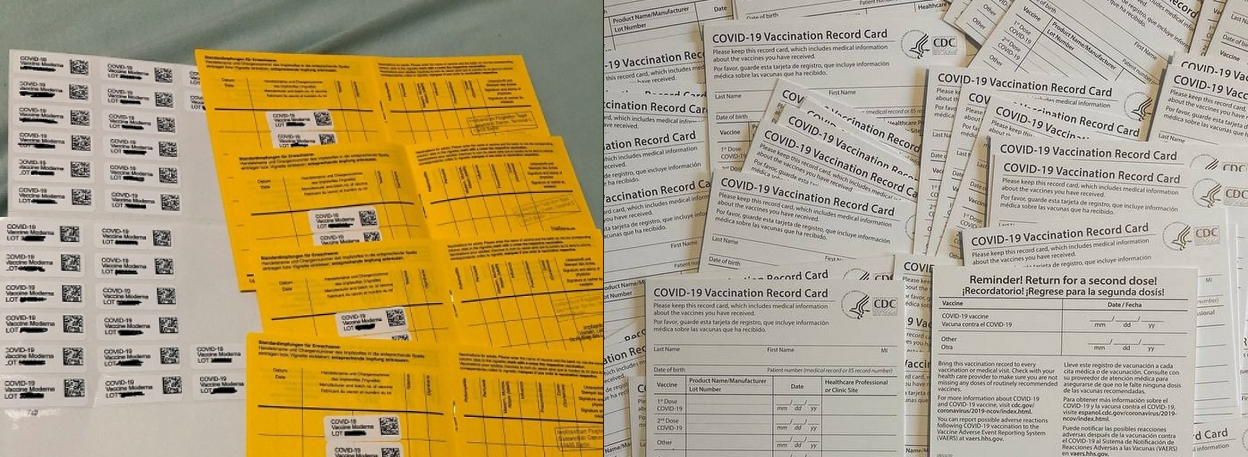}
            \vspace{-0cm}\caption{Left: Verified seller advertising full German vaccination certificates. Right: Physical USA certificate example as advertised on a darkweb marketplace.}
            \label{fig:papercerts}
            \end{figure}

    While looking for certificate listings on the darkweb we found that a number of platforms forbid vendors from listing COVID-19 related items. 
    The ones that allowed for such listings, in most cases included both physical (e.g. U.S. vaccination cards, see Figure \ref{fig:papercerts}) and digital certificates (e.g. EU green certificate). We documented listings from Germany and the Netherlands, which along with the QR code, also contained a ''yellow vaccination card'' \cite{germancert, dutchcert} with a sticker tag of the vaccine's ID number, accompanied by a vaccination center's stamp and a doctor's signature (see Figure \ref{fig:papercerts}). This card alone can be used at a national level, but it cannot serve as a proof of vaccination in other EU countries (with minor exceptions). Furthermore, there were also listings offering vaccinated individual's QR codes, instead of generating a new one for each specific customer, and mentioned that in order to use these QR codes, the buyers should also acquire a fake ID under the same name.

\begin{figure}
            \centering
            \includegraphics[width=0.99\columnwidth]{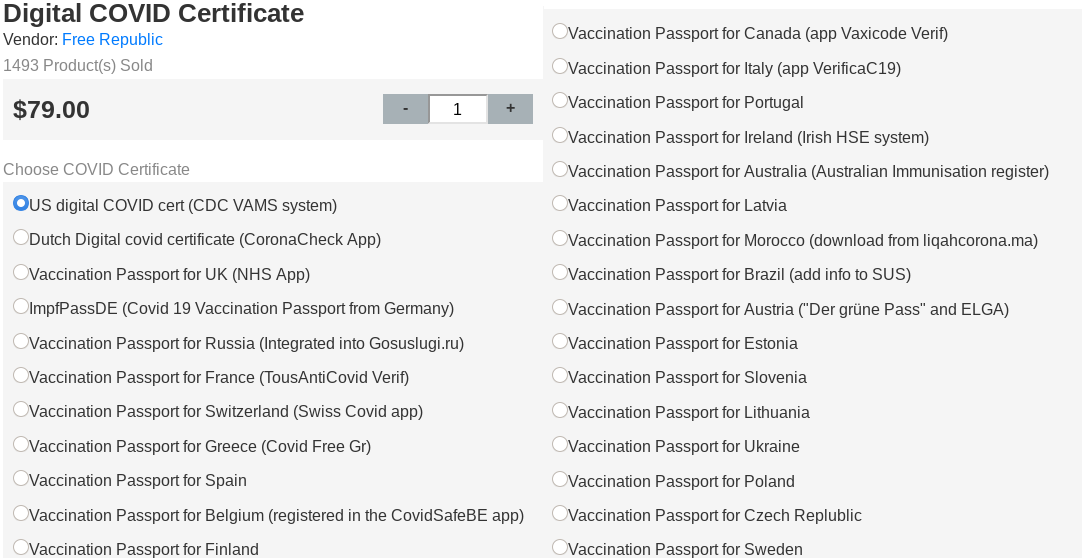}
            \vspace{-0cm}\caption{Example of registration countries for COVID-19 vaccination certificates.}
            \label{fig:countries}
            \end{figure}

    Regarding the registration countries, the listings are heavily focused on European countries, and the United States, but there are also listings from other continents and countries, such as Brazil, Canada, Mexico and Australia (see Figure \ref{fig:countries}). Destination countries presented more variety and flexibility, with many listings mentioning worldwide shipping.

\subsection{Promotion}
    Every successful business, needs to advertise their services and products in some way. In the case of vaccination certificates, a very effective way to achieve this is by utilizing darkweb forums. Forum discussions on the COVID-19 global situation mostly include members debating on whether the vaccines are safe to use, or commenting on conspiracy theories. In addition, there are also posts from members sharing their own story and reasons behind their need to acquire a forged certificate, often both for themselves and their family. These discussions, are a good option for vendors to advertise their services, with many posts providing links to marketplaces and vendor shops, or specific information on how to acquire a certificate. Moreover, other forum users would also refer the buyers to vendors and platforms. Lastly, vendors are also trying to attract clients using specific phrases in their listings in regards to the vaccines, such as ''\textit{poison}'', ''\textit{health of you and your loved ones}'', ''\textit{danger}'' and ''\textit{freedom}''.

 \subsection{Communication}
     For the listings found on marketplaces, communication with the sellers takes place through the integrated messaging function of the platform, since vendors are forbidden from contacting clients privately. On the contrary, certificate listings on vendor shops, apart from the platform's own messaging system, also included other methods of communication, such as \textit{ProtonMail}, \textit{Wickr}, and \textit{Telegram}.
     
     \subsection{Vaccine options}
     The majority of certificates available are of vaccination with the \textit{Pfizer-BioNTech} vaccine. Nevertheless, there are also listings of the \textit{Moderna}, \textit{Johnson \& Johnson}, as well as the Russian \textit{Sputnik-V} vaccine, with the latter specifically found in the \textit{Hydra} marketplace.

\subsection{Payment methods \& currency}
\label{payment}
\begin{figure}
            \centering
            \includegraphics[width=0.4\columnwidth]{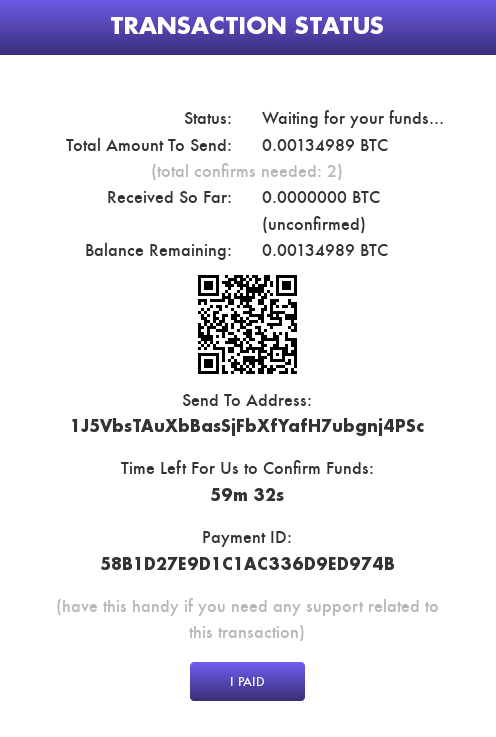}
            \vspace{-0cm}\caption{Payment example of a COVID-19-specific vendor shop.}
            \label{fig:payment}
            \end{figure}

    Depending on the platform that the listings are available on, the possible payment methods differ. 
    On vendor shops, the accepted payment method is direct payment. In this scenario, the vendor provides a cryptocurrency address which the buyer can then use to deposit the funds for the purchase. An example of a vendor shop payment option is depicted in Figure \ref{fig:payment}. 
    
    In the case of marketplaces, the payment method is through the \textit{escrow} mechanism \cite{soska2015measuring}. According to this mechanism, the marketplace provides a deposit address to the client to deposit the funds, and holds onto them until the client verifies that the order has been received and everything is as expected. This provides much more assurance to the client that they will not get ''scammed'' by dishonest vendors, also known as \textit{rippers}. Furthermore, \textit{Multisignature Escrow} \cite{soska2015measuring} is not utilized by any of the marketplace vendors.
    
    The pricing differs greatly between the different listings, with the cheapest certificate starting at \$39 and the highest price reaching almost \$2,800, which included both a physical and a digital certificate, registered in the United Kingdom.

    In terms of accepted currency, the majority of vendor shops are using \ac{BTC}, while marketplaces also include \ac{XMR}. However, there are some special cases where more cryptocurrencies are accepted, such as \ac{ETH}, \ac{ADA}, \ac{LTC} and \ac{ZEC}.

\subsection{Purchase \& delivery}
    On marketplaces, the purchase process includes adding the product to the cart (similarly to legitimate online shops) and then either using a cryptocurrency address (generated for the specific purchase) provided by the marketplace to deposit the funds from their off-site wallet, or paying using the account balance (on site-wallet). In both scenarios, the transaction is carried out through the \textit{escrow} mechanism (see Section \ref{payment}). 
    After the payment has been confirmed by the marketplace, the client needs to provide their private information. This varies depending on the country that the certificate is being issued in and may include full name, country, address, social security number, state, zip code, and email address\footnote{This provides the opportunity to scammers to not only steal people's money but also their private information (see also Section \ref{scammers}).}. Such information, depending on the platform, is sent via different methods, with the main ones being \textit{ProtonMail} and the platforms' own messaging functions (on marketplaces, this information is also encrypted with \ac{PGP}, using the seller's public key). All of this information is then used by the sellers to create the valid COVID-19 certificate. This process also applies to the case of vendor shops, with the difference that, as discussed above (see Section \ref{payment}), there is no escrow mechanism in place, and the payment is direct.

    After purchasing a certificate, the delivery method depends on whether it is in physical form or in digital. For vaccination cards (physical), the shipping method advertised is normal post. However, in the case of digital certificates there is more variety.
    Several vendors advertise that after issuing the certificate, they will notify the buyer via email. The buyer would then use the national mobile COVID-19 application to use the certificate, as if they were vaccinated. In other cases, the sellers claim that they can send the vaccination confirmation via email, using the official email address of the hospital that they accessed the issuer service from (see Section \ref{background}). After the confirmation, the buyer also receives the certificate PDF file.

%% file: Chapters/5.Valid.tex
\section{Legitimacy and use cases}
\label{working}
This section is dedicated to investigating the legitimacy of the certificate listings we discovered on the darkweb platforms. 
\subsection{Scammers}
\label{scammers}
\begin{figure}
            \centering
            \includegraphics[width=0.7\columnwidth]{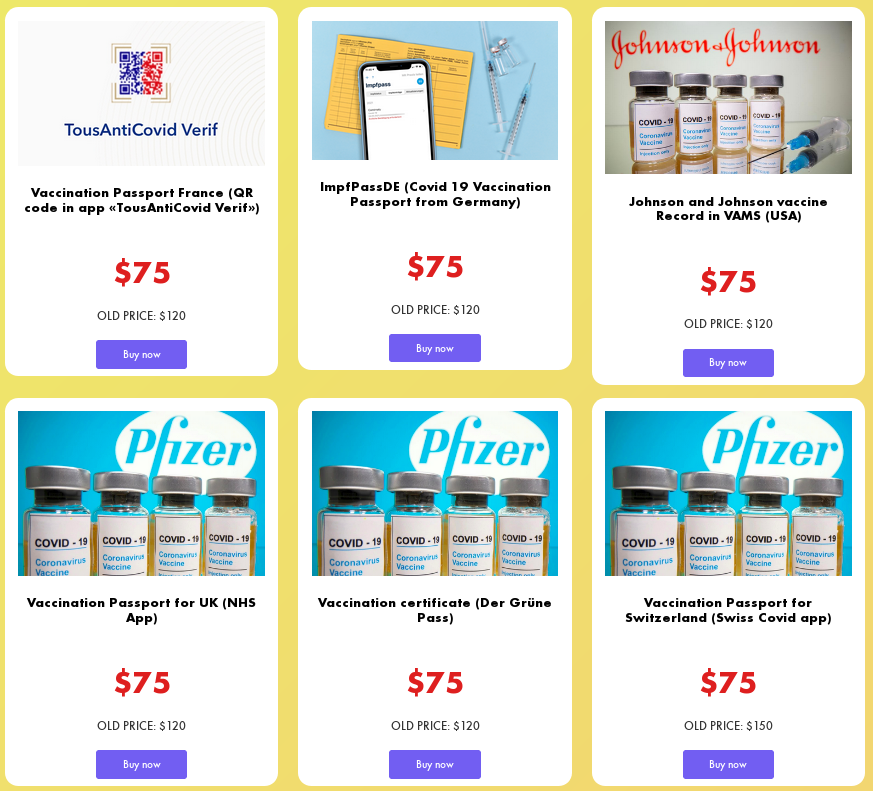}
            \vspace{-0cm}\caption{Vaccination certificate options of a COVID-19-specific vendor shop. }
            \label{fig:options}
            \end{figure}

   While investigating certificate selling platforms, we noticed various indications that can point towards a vendor being a scammer, with extreme prices, both too low and too high, being one of them. Many platforms also provided additional types of products, such as vaccines, with all of them offered at the same base price (see Figure \ref{fig:options}). Another indication was the absence of sale history, vendor ratings, reviews, or any type of feedback. One element that could serve as a potential motivation for clients to trust a specific vendor, is the payment method. All of the vendor shops we explored, use direct payments, while marketplaces implemented the escrow payment mechanism (see Section \ref{payment}). With such a mechanism in place, it is less likely for a client to be scammed, since the marketplace will not release the funds to the vendor unless the customer verifies that their order was delivered. Likewise, vendors are less likely to advertise services they cannot actually provide.


    Another method to lure clients, with the intention of deceiving them is using stolen valid certificates. One of the marketplaces we explored, included a listing offering digital certificates at the price of \$405, that as mentioned by the sellers, have the same validity, as if the vaccination had taken place in France. Additionally, the listing also included a screenshot of a certificate, which we analyzed using two different mobile COVID-19 applications (see Section \ref{method}). The certificate turned out to be valid. The results pointed to a real person, based on the social media activity we stumbled upon by carrying out a simple web search, who we managed to get in touch with through a popular social media platform. After explaining the context of our finding, the owner informed us that they had publicly showcased the certificate for professional reasons. The vendor that created the specific listing on the marketplace was a very new vendor (ten day old account), had no sales, no feedback from other users, and in general no activity of any type that could serve as an indication of their legitimacy. For that reason, it seems that the vendor simply used the valid certificate as an assurance to the potential clients that the advertised service can be delivered, in an effort to jump start their business.

\subsection{Verified sellers}
    \subsubsection{Indirect verification}
    Even though, as mentioned, the majority of certificate listings did not include reviews or any sale history, there were some exceptions. Some marketplace vendors, present previous sales of the actual certificate product they are advertising, or sales of other products, which result in a higher trust status on the platform. Additionally, these vendors have client reviews from past buyers, and also support payments through escrow (see Section \ref{payment}). The pricing on these listings fluctuated to a significant extent, with some starting from \$39 and reaching \$2,000, and did not correlate to the number of positive reviews, or the number of total sales.

    \subsubsection{Direct verification}

\begin{figure}
            \centering
            \includegraphics[width=0.6\columnwidth]{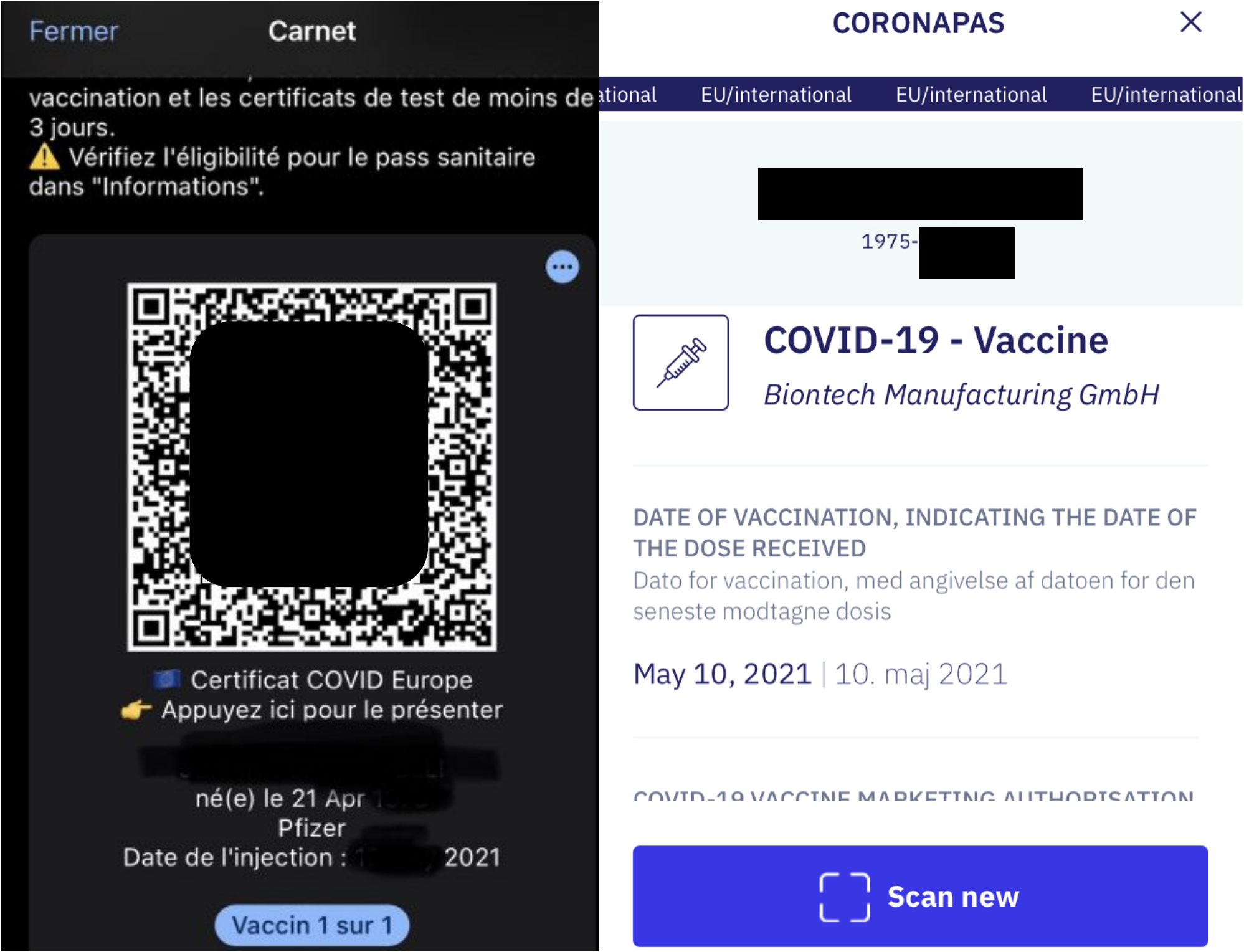}
            \vspace{-0cm}\caption{The valid certificate. Left: the QR-code of the certificate (anonymized for the purposes of this paper). Right: Our validation of the certificate using the Danish national mobile application ''Coronapas''.}
          \label{fig:joanna}
            \end{figure}

         \paragraph{A valid certificate for an non-identifiable person}
        Among the platforms we investigated, there was a specific vendor shop advertising COVID-19 certificates registered in France, for the price of \$450. The listing included no reviews, no feedback, no sale history, or any type of buyer feedback whatsoever. It only presented a screenshot taken from the official COVID-19 mobile application of France, illustrating a QR code, a date of birth, and a date of vaccination. We decided to use the Danish and French COVID-19 national mobile applications, to test the validity of the QR code (see Figure \ref{fig:joanna}). Based on our findings, the QR code is actually valid and operational, giving a result like any other QR would, in the case of a legitimately vaccinated individual (signed by the ``\textit{Caisse nationale de l'assurance maladie (CNAM)})''. 
        
        We attempted to validate the existence of the person the certificate was issued for in an effort to validate the results. After conducting many web searches, the only result coming up was that of a woman of German ancestry, dating back to the 18th century. We note that the date of birth stated on the certificate, is very similar to the date of birth of the woman from the $1700$s. Specifically, the day is an exact match, the month number is increased by one, and the year is changed in a way that the last two digits match the exact age of the said woman, at her time of death. However, since we have no way to verify the meaning of this information, it is just conjecture. Lastly, the certificate also includes the exact number of the vaccine doses that have been administered, along with the manufacturer's name, \textit{BioNTech, Pfizer}. Specifically, the certificate states that \textit{''Doses 1 of 1''} have been administered, which is very out of the ordinary since it is well known that the Pfizer vaccine requires two doses for the vaccination process to be finalized \cite{doses}. This fact can be an indication that the certificate has been forged, while maintaining validity.
        
        \paragraph{A vendor shop with advanced forging capabilities}
        Out of the entire certificate trading market, there was a particular instance that really stood out. This instance refers to a particular vendor shop, which is the only platform that elaborates on the operation of their service in such detail. On this site, the vendors illustrate technical information on the architecture of the issuance and verification mechanisms in the EU, as well as on the specifications of the JSON token of the certificate QR code \cite{jsonspecs}. The information they provide are in conformity with the official European Commission specifications, as described in Section \ref{background} and depicted on Figures \ref{fig:architecture} and \ref{fig:json}. Both figures have been developed by the vendors and are presented on their platform.
        
        The sellers advertise certificates registered in $25$ EU countries and also present reviews from past buyers, along with the overall amount of sales they have carried out. The price for a single certificate is at \texteuro250, with bundles of two or three certificates available at \texteuro400 and \texteuro525 respectively. All payments are exclusively carried out using \ac{BTC}. To provide proof that the generated certificates sold are valid, the homepage of the site also includes a sample QR code, of a fictional individual, which we validated using two national COVID-19 mobile applications. With regard to communication, the methods of choice of this vendor shop include \textit{ProtonMail} and \textit{Telegram}, but there is also an integrated form functionality intended for customers that have already paid and received their order token. This token is a requirement in order to submit a ticket through the form.

        While trying to acquire additional information on the vendor shop, we also discovered an unlisted \textit{YouTube} video, which contained a lot of extra detail regarding the platform's operation. Clients can choose all of the widely used vaccines in the EU, like \textit{Pfizer} and \textit{Moderna}, however the vendors themselves recommend choosing the Pfizer vaccine, since it provides more ease and speed in regards to the delivery of the certificate. Additionally, they state that in the case of an additional third or second vaccine ''booster'' shot being necessary in the future (depending on the vaccine type), they will update the certificates that have already been sold, free of charge. The vendors also demonstrate the payment mechanism of the site, according to which the clients need to firstly pay for the certificate, send their information after the payment's verification, and then wait for the vendors to provide the QR code. In this video, the viewer can also catch a short glimpse of the service administration dashboard, where the interface used to generate the certificates is clearly visible. Additionally, the total amount of sales, along the overall revenue generated, can both also be seen. Specifically, the platform seems to have made over $1700$ sales, which amounts to more than \$450,000 in profit.

\begin{figure}
            \centering
            \includegraphics[width=0.6\columnwidth]{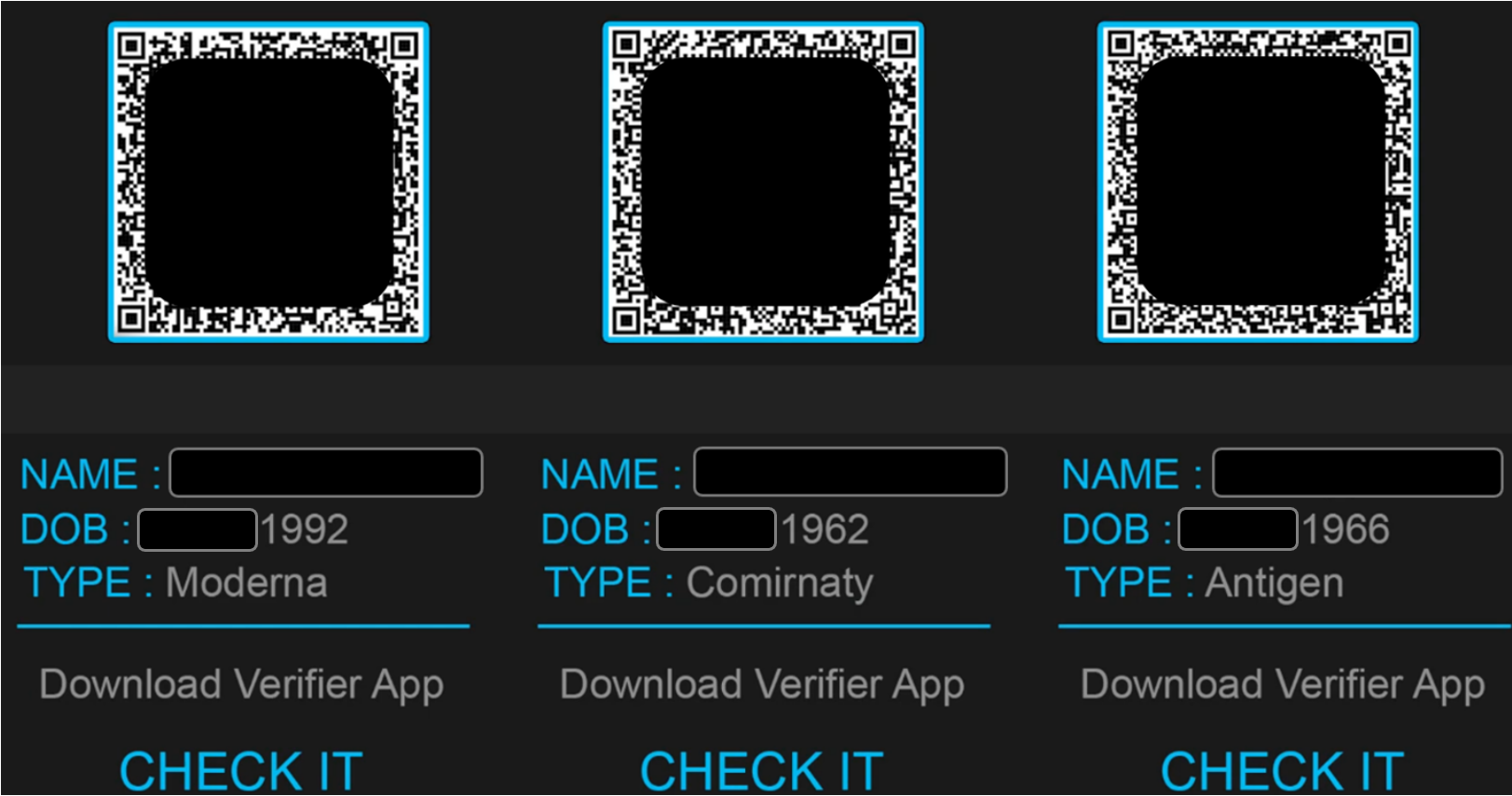}
            \vspace{-0cm}\caption{Vendor demonstrating synthetic QR code generation capabilities for all type of vaccines. All of these have been checked and appear to be valid when scanned.}
          \label{fig:greenpass}
            \end{figure}

        Furthermore, the vendors showcase the verification of a great number of certificates that they have generated for demonstration purposes, with all of them containing credentials which, according to the vendors, do not correspond to any real person. All of these certificates seem to have been registered in Italy. The demonstration process is carried out through a mobile application developed by the vendors, so in order to be certain that the certificates are indeed valid, and this is not some type of scam, we yet again verified the certificates using the two national mobile COVID-19 applications. We did not manage to verify all of them due to issues in capturing a high quality frame in the video, but the ones we did manage to check, appear to be valid. 
        Additionally, in the video the sellers provide three specific QR codes for verification purposes, which also turned out to be valid (see Figure \ref{fig:greenpass}).
        They also mention that they prefer smaller private channels to advertise their business, to avoid unwanted attention and not allow for scam opportunities (e.g. development of phishing sites). Lastly, the vendors stress the fact that they do not keep any of the personal information provided to them by the clients, since they have no intention of monetizing this data, and want to provide their services without jeopardizing their customers' privacy.

        The individuals behind this vendor shop present an advanced understanding of the system that surrounds the issuance and verification of certificates, which combined with the quality of their web page, the overall attention to detail in describing the operation of their business, and the verification use cases shown, raises the probability of the service being legitimate. This fact however, leads to the question of how these sellers have managed to infiltrate the EU COVID-19 certificate systems in so many countries. Unfortunately, they do not disclose this information, since it would mean the end of their operation.

    \subsection{Certificate supply}
    \label{supply}
    At this point, the question that rises is how these individuals are able to acquire valid COVID-19 certificates. To assure the potential clients, that their product is valid and usable in a real life scenario, many vendors include information in the description of the product, explaining how they are able to provide such a service. The majority of sellers claim having access to the system tasked with the issuance of certificates, such as the EU issuer web application (see Section \ref{background}). The access to these systems is allegedly achieved through individuals working in hospital or health organizations, who are able to access the issuance mechanisms, or by remotely hacking them. 
    
    In the specific case of a listing on the Russian marketplace \textit{Hydra}, the description even mentioned the exact location and hospital that the system was accessed from. Although not advertised on any of the listings we discovered, an additional way that would allow for the issuance of valid certificates, is by somehow acquiring the private key (or keys) of an issuance body (e.g. health organization), with a recent case of forged COVID-19 certificates surfacing, suggesting as such \cite{fakecerts}. This would provide the ability to generate an unlimited number of valid certificates at will. 
    
    We want to note that if the latter case (leaked keys) is true there is a problem with the current EU certificate model. While there is the option for EU countries to mark a key as compromised or invalid, and revoke the corresponding certificate \cite{revocation}, doing so would automatically mark a large amount of benign certificates invalid as well. 

%% file: Chapters/6.Conclusion.tex
\section{Conclusion}

Vendors on the darkweb were very quick in attempting to exploit the worldwide need for COVID-19 vaccines and certificates. They have managed to create a new market, which initially preyed on people's fear of getting infected with the virus, by offering vaccines even before they were officially rolled out by the governmental entities. This market has now manifested into a vaccination certificate trading market, this time founded on the lack of trust in the vaccine by a part of the global population, as well as the resentment of these individuals towards the everyday life restrictions and hardships that come as a result of not being vaccinated. 

By investigating $27$ darkweb COVID-19 certificate selling platforms, we acquire insight on this specific market regarding its products and properties. Out of all the listings we found, it is very challenging to verify how many of them refer to valid certificates. This is mainly due to the fact that the market is not as established yet, as others on the darkweb, with the drug trade as the primary example. This state leads to a few sales, lack of client feedback, and the majority of sellers being unverified. However, we manage to discover a number of certificates which we are able to verify, raising the issue of malicious individuals having access to governmental systems, which they can manipulate at will, or keys of national health organizations' having leaked. Specifically, in one of the cases discussed, the vendors present a worrisome level of efficiency and competence. We hope that this article will raise awareness on the situation, motivating the corresponding authorities to further investigate the security of the current certificate issuance systems.

%% file: Main.bbl
\begin{thebibliography}{10}

\bibitem{uscerts}
Harini~Barath Bobbie~Johnson, Adriana~Fraser.
\newblock What’s happening with covid vaccine apps in the us.
\newblock
  \url{https://www.technologyreview.com/2021/08/31/1033993/vaccine-credential-initiative-us-state-guide/}.

\bibitem{bracci2021dark}
Alberto Bracci, Matthieu Nadini, Maxwell Aliapoulios, Ian Gray, Damon McCoy,
  Alexander Teytelboym, Angela Gallo, and Andrea Baronchelli.
\newblock Dark web marketplaces and covid-19: The vaccines.
\newblock {\em Available at SSRN 3783216}, 2021.

\bibitem{bracci2}
Alberto Bracci, Matthieu Nadini, Maxwell Aliapoulios, Damon McCoy, Ian Gray,
  Alexander Teytelboym, Angela Gallo, and Andrea Baronchelli.
\newblock Dark web marketplaces and covid-19: before the vaccine.
\newblock {\em EPJ data science}, 10(1):6, 2021.

\bibitem{christin2013traveling}
Nicolas Christin.
\newblock Traveling the silk road: A measurement analysis of a large anonymous
  online marketplace.
\newblock In {\em Proceedings of the 22nd international conference on World
  Wide Web}, pages 213--224, 2013.

\bibitem{zlib}
L.~Peter Deutsch and Jean-Loup Gailly.
\newblock Zlib compressed data format specification version 3.3.
\newblock \url{https://datatracker.ietf.org/doc/html/rfc1950}.

\bibitem{cose}
August~Cellars Jim~Schaad.
\newblock Cbor object signing and encryption (cose).
\newblock \url{https://datatracker.ietf.org/doc/html/rfc8152}.

\bibitem{cbortoken}
Michael~B. Jones, Erik Wahlström, Samuel Erdtman, and Hannes Tschofenig.
\newblock Cbor web token (cwt).
\newblock
  \url{https://tools.ietf.org/id/draft-ietf-ace-cbor-web-token-15.html}.

\bibitem{9558786}
Georgios Karopoulos, Jose~L. Hernandez-Ramos, Vasileios Kouliaridis, and
  Georgios Kambourakis.
\newblock A survey on digital certificates approaches for the covid-19
  pandemic.
\newblock {\em IEEE Access}, pages 1--1, 2021.

\bibitem{russiareuters}
Andrey Ostroukh and Gabrielle Tétrault-Farber.
\newblock Russian regions introduce qr codes for entry to public venues as
  covid-19 cases hit record.
\newblock
  \url{https://www.reuters.com/world/europe/russian-regions-introduce-qr-codes-entry-public-venues-covid-19-cases-hit-record-2021-10-18/}.

\bibitem{doses}
\relax Centers~for Disease~Control and Prevention.
\newblock Covid-19 vaccines that require 2 shots.
\newblock
  \url{https://www.cdc.gov/coronavirus/2019-ncov/vaccines/second-shot.html}.

\bibitem{jsonspecs}
\relax European~Commission.
\newblock Technical specifications for \uppercase {EU} digital covid
  certificates json schema specification.
\newblock
  \url{https://ec.europa.eu/health/sites/default/files/ehealth/docs/covid-certificate\_json\_specification\_en.pdf}.

\bibitem{europacerts}
\relax {European}~Commission.
\newblock \uppercase {EU} digital covid certificate.
\newblock
  \url{https://ec.europa.eu/info/live-work-travel-eu/coronavirus-response/safe-covid-19-vaccines-europeans/eu-digital-covid-certificate\_en}.

\bibitem{cscas}
\relax European Commission~eHealth Network.
\newblock Technical specifications for digital green certificates volume 1.
\newblock
  \url{https://ec.europa.eu/health/sites/default/files/ehealth/docs/digital-green-certificates\_v1\_en.pdf}.

\bibitem{certtechspecs}
\relax European Commission~eHealth Network.
\newblock Technical specifications for digital green certificates volume 4.
\newblock
  \url{https://ec.europa.eu/health/sites/default/files/ehealth/docs/digital-green-certificates\_v4\_en.pdf}.

\bibitem{revocation}
\relax European Commission~eHealth Network.
\newblock Technical specifications for digital green certificates volume 5.
\newblock
  \url{https://ec.europa.eu/health/sites/default/files/ehealth/docs/digital-green-certificates\_v5\_en.pdf}.

\bibitem{dutchcert}
\relax The Hague~Online.
\newblock The usefulness of a covid stamp in your yellow booklet: read here how
  it works.
\newblock
  \url{https://www.thehagueonline.com/news/2021/06/14/the-usefulness-of-a-covid-stamp-in-your-yellow-booklet-read-here-how-it-works}.

\bibitem{germancert}
Lisa Schreiner.
\newblock What you need to know about germany’s “impfpass” vaccination
  record.
\newblock
  \url{https://www.iamexpat.de/expat-info/german-expat-news/what-you-need-know-about-germanys-impfpass-vaccination-record}.

\bibitem{farmer}
Mathew~J. Schwartz.
\newblock Feds bust 'farmer's market' for online drugs.

\bibitem{soska2015measuring}
Kyle Soska and Nicolas Christin.
\newblock Measuring the longitudinal evolution of the online anonymous
  marketplace ecosystem.
\newblock In {\em 24th $\{$USENIX$\}$ security symposium ($\{$USENIX$\}$
  security 15)}, pages 33--48, 2015.

\bibitem{fakecerts}
Lisa Vaas.
\newblock Update: Eu’s green pass vaccination id private key leaked or
  forged.
\newblock
  \url{https://threatpost.com/eus-green-pass-vaccination-id-private-key-leaked/175857/}.

\bibitem{vu2020turning}
Anh~V Vu, Jack Hughes, Ildiko Pete, Ben Collier, Yi~Ting Chua, Ilia Shumailov,
  and Alice Hutchings.
\newblock Turning up the dial: the evolution of a cybercrime market through
  set-up, stable, and covid-19 eras.
\newblock In {\em Proceedings of the ACM Internet Measurement Conference},
  pages 551--566, 2020.

\end{thebibliography}
